\documentclass[prd,aps,nofootinbib,floatfix,11pt]{revtex4}
\usepackage{amsmath,graphicx,epsfig,amssymb,dsfont,mathtools,}
\usepackage[usenames]{color}
\usepackage{ulem} %% for strike-through
\usepackage{bigstrut}
\usepackage{slashed}
\usepackage{multirow}

\allowdisplaybreaks

\begin{document}\title{Light-Cone Sum Rules Analysis of $\Xi_{QQ^{\prime}q}\to\Lambda_{Q^{\prime}}$ Weak Decays}
\author{Yu-Ji Shi$^{1}$~\footnote{Email:shiyuji@sjtu.edu.cn}, Ye Xing$^{1}$~\footnote{Email:xingye guang@sjtu.edu.cn},
and Zhen-Xing Zhao$^{1}$~\footnote{Email:star\_0027@sjtu.edu.cn}}
\affiliation{$^{1}$ INPAC, SKLPPC,
School of Physics and Astronomy, Shanghai Jiao Tong University, Shanghai 200240, China}
\begin{abstract}
We analyze the weak decay of doubly-heavy baryon decays into anti-triplets $\Lambda_Q$ with light-cone sum rules. To calculate the decay form factors, both bottom and charmed anti-triplets $\Lambda_b$ and $\Lambda_c$ are described by the same set of leading twist light-cone distribution functions. With the obtained form factors, we perform a phenomenology study on the corresponding semi-leptonic decays. The decay widths are calculated and the branching ratios given in this work are expected to be tested by future experimental data, which will help us to understand the underlying dynamics in doubly-heavy baryon decays. 
\end{abstract}
\maketitle

%%%%%%%%%%%%%%%%%%%%%
\section{Introduction}

Since the establishment of the quark model, people have attempted to construct a complete hadron spectrum containing all the particles predicted by the model. Although in the past few decades lots of hadron states have been observed from experiments, there still remains some predicted but unobserved particles, even in their ground states. One kind of such particles is doubly-heavy baryon, which consists of two heavy flavor quarks and a light quark. In 2017, the LHCb collaboration announced the observation of the ground state doubly-charmed baryon $\Xi_{cc}^{++}$ which has the mass~\cite{Aaij:2017ueg}
\begin{equation}
m_{\Xi_{cc}^{++}}=(3621.40\pm0.72\pm0.27\pm0.14)\ {\rm MeV}.\label{eq:LHCb_measurement}
\end{equation}
This newly observed particle was reconstructed from the decay channel $\Lambda_{c}^{+}K^{-}\pi^{+}\pi^{+}$,  which had been predicted in Ref.~\cite{Yu:2017zst}. Only a year later LHCb announced their measurement on $\Xi_{cc}^{++}$ lifetime~\cite{Aaij:2018wzf} as well as observation on a new two-body decay channel $\Xi_{cc}^{++}\to\Xi_{c}^{+}\pi^{+}$~\cite{Aaij:2018gfl}. Recently, experimentalists are continuing to search for other heavier particles included in the doubly-heavy baryon spectroscopy~\cite{Traill:2017zbs,Cerri:2018ypt}. On the other hand, the great progress on the experiments also make the study of doubly-heavy baryons become a hot topic of theoretic high energy physics. Up to now there have been many corresponding theoretic studies which aim to understand the dynamic and spectroscopy properties of the doubly-heavy baryon states~\cite{Wang:2017mqp,Meng:2017udf,Wang:2017azm,
Gutsche:2017hux,Li:2017pxa,Guo:2017vcf,Xiao:2017udy,Sharma:2017txj,Ma:2017nik,Hu:2017dzi,Shi:2017dto,Yao:2018zze,Yao:2018ifh,
Ozdem:2018uue,Ali:2018ifm,Dias:2018qhp,Zhao:2018mrg,Xing:2018bqt,
Ali:2018xfq,Liu:2018euh,Xing:2018lre,Bediaga:2018lhg,Wang:2017vnc,Dhir:2018twm,Berezhnoy:2018bde,Jiang:2018oak,Zhang:2018llc,Li:2018bkh,Gutsche:2018msz}.

Semi-leptonic doubly-heavy baryon weak decay offers an ideal platform for studying such baryon states. The main advantage is that the weak and strong dynamics are separated in semi-leptonic processes, while the QCD effects are totally capsuled in the hadron transition matrix element, which is parametrized by six form factors. In the literature, there are some results of calculating doubly-heavy baryon form factors with light-front quark model (LFQM)~\cite{Wang:2017mqp,Zhao:2018mrg}. In a previous work, we derived these form factors with QCD sum rules (QCDSR)~\cite{Shi:2019hbf}. We performed a leading order calculation for a three-point correlation function by OPE, where the contribution of the local operators ranging from dimension 3 to 5 are summed. In this work, we will perform a calculation for doubly-heavy baryon form factors with light-cone sum rules (LCSR). In the framework of LCSR, one uses non-local light-cone expansion instead of the local OPE, while the non-perturbative effect is produced by light-cone distribution amplitudes (LCDAs) of hadron instead of the vacuum condensates.  When using LCSR for studying form factors, one only needs a two-point correlation function for calculation. The great advantage of this is not only that the two-point correlation function is much easier to be dealt with, but also it avoids the potential irregularities of the truncated OPE in the three-point sum rules~\cite{Colangelo:2000dp}. 

In this work we will use LCSR to study $\Xi_{cc}$, $\Xi_{bb}$ or $\Xi_{bc}$ baryon weak decays and the final state baryon is focused on an anti-triplet $\Lambda_b$ or $\Lambda_c$. The quark level transition can be either $b\to u$ or $c\to d$. This paper is arranged as follows. In Sec.~\ref{sec:lc_sum_rules}, we will introduce the definition of the transition form factors of doubly heavy baryon weak decays. Then with the introduction of the light-cone distribution amplitudes of $\Lambda_Q$ baryons, we will illustrate the LCSR approach for deriving the transition form factors. In Sec.~\ref{sec:numerical results}, we will give the numerical results for the form factors and use them to calculate decay widths as well as branching ratios of doubly heavy baryon semi-leptonic decays. Sec.~\ref{sec:conclusions} is a summary of this work and the prospect of LCSR study on doubly-heavy baryons for the future.

\section{Transition Form Factors in light-cone sum rules}
\label{sec:lc_sum_rules}
%%%%%%%%%%%%%%%%%%%

\subsection{Form Factors}

To parametrize the hadron transition $\Xi_{QQ^{\prime}q}\to\Lambda_{Q^{\prime}}$, six form factors are defined: 
\begin{eqnarray}
&&\langle{\Lambda_{Q^{\prime}}}(p_{\Lambda},s_{\Lambda})|(V-A)_{\mu}|{\Xi_{QQ^{\prime}q}}(p_{\Xi},s_{\Xi})\rangle \nonumber \\ & = & \bar{u}_{\Lambda}(p_{\Lambda},s_{\Lambda})\bigg[\gamma_{\mu}f_{1}(q^{2})+i\sigma_{\mu\nu}\frac{q^{\nu}}{m_{\Xi}}f_{2}(q^{2})+\frac{q_{\mu}}{m_{\Xi}}f_{3}(q^{2})\bigg]u_{\Xi}(p_{\Xi},s_{\Xi})\nonumber \\
&  &- \bar{u}_{\Lambda}(p_{\Lambda},s_{\Lambda})\bigg[\gamma_{\mu}g_{1}(q^{2})+i\sigma_{\mu\nu}\frac{q^{\nu}}{m_{\Xi}}g_{2}(q^{2})+\frac{q_{\mu}}{m_{\Xi}}g_{3}(q^{2})\bigg]\gamma_{5}u_{\Xi}(p_{\Xi},s_{\Xi}), 
\label{eq:parameterization1}
\end{eqnarray}
The (spinor, momentum, mass, helicity) of the initial and the final baryons are ($u_{\Xi},\ p_{\Xi},\ m_{\Xi},\ s_{\Xi}$) and ($u_{\Lambda},\ p_{\Lambda},\ m_{\Lambda},\ s_{\Xi}$) respectively. The weak decay current is composed by a vector current $\bar q \gamma^\mu Q$ and a axial-vector current $\bar q \gamma^\mu\gamma^5 Q$, where $q$ denote a light quark while $Q$ denote a bottom or charm quark. $f_i(q^2)$ and $g_i(q^2)$ are two sets of form factors parametrizing the vector current induced and the axial-vector current induced transitions respectively. The transfering momentum is defined as $q^\mu= p_{\Xi}^\mu -p_{\Lambda}^\mu$.

To simplify the calculations, one can also use the following parametrizing convention
\begin{eqnarray}
&&\langle{\Lambda_{Q^{\prime}}}(p_{\Lambda},s_{\Lambda})|(V-A)_{\mu}|{\Xi_{QQ^{\prime}q}}(p_{\Xi},s_{\Xi})\rangle \nonumber \\ & = & \bar{u}_{\Lambda}(p_{\Lambda},s_{\Lambda})\bigg[F_{1}(q^{2})\gamma_{\mu}+F_{2}(q^{2})p_{\Lambda}+F_{3}(q^{2})p_{\Xi}\bigg]u_{\Xi}(p_{\Xi},s_{\Xi})\nonumber \\
&  &- \bar{u}_{\Lambda}(p_{\Lambda},s_{\Lambda})\bigg[G_{1}(q^{2})\gamma_{\mu}+G_{2}(q^{2})p_{\Lambda}+G_{3}(q^{2})p_{\Xi}\bigg]\gamma_{5}u_{\Xi}(p_{\Xi},s_{\Xi}), 
\label{eq:parameterization2}
\end{eqnarray}
Such definition enables us to simply extract the $F_i$ and $G_i$ in the frame work of LCSR. These form factors are related with those defined in Eq. (\ref{eq:parameterization2}) as
\begin{eqnarray}
f_1(q^2) &=& F_1(q^2)+\frac{1}{2}(m_{\Xi}+m_{\Lambda})(F_2(q^2)+F_3(q^2)), \nonumber \\
f_2(q^2) &=& \frac{1}{2}m_{\Xi}(F_2(q^2)+F_3(q^2)), \nonumber \\
f_3(q^2) &=& \frac{1}{2}m_{\Xi}(F_3(q^2)-F_2(q^2)),
\end{eqnarray}
\begin{eqnarray}
g_1(q^2) &=& G_1(q^2)-\frac{1}{2}(m_{\Xi}-m_{\Lambda})(F_2(q^2)+F_3(q^2)), \nonumber \\
g_2(q^2) &=& \frac{1}{2}m_{\Xi}(G_2(q^2)+G_3(q^2)), \nonumber \\
g_3(q^2) &=& \frac{1}{2}m_{\Xi}(G_3(q^2)-G_2(q^2)).
\end{eqnarray}

\subsection{Light-Cone Distribution Amplitudes of $\Lambda_{Q}$}

The light-cone distribution functions of singly-heavy baryons were derived in Ref.~\cite{Ball:2008fw,Ali:2012pn} by the approach of QCDSR at the heavy quark mass limit. In this work we use the LCDAs of $\Lambda_{b}$ from Ref.~\cite{Ball:2008fw}, which are defined by the following four matrix elements of nonlocal operators:
\begin{eqnarray}
\frac{1}{v_{+}}\langle0|[q_{1}^{T}(t_{1})C\gamma_{5}\slashed nq_{2}(t_{2})]Q_{\gamma}(0)|\Lambda_{Q}(v)\rangle & =\psi_{2}(t_{1},t_{2})f^{(1)}u_{\gamma},\nonumber\\
\frac{i}{2}\langle0|[q_{1}^{T}(t_{1})C\gamma_{5}\sigma_{\mu\nu}q_{2}(t_{2})]Q_{\gamma}(0)\bar{n}^{\mu}n^{\nu}|\Lambda_{Q}(v)\rangle & =\psi_{3\sigma}(t_{1},t_{2})f^{(2)}u_{\gamma},\nonumber\\
\langle0|[q_{1}^{T}(t_{1})C\gamma_{5}q_{2}(t_{2})]Q_{\gamma}(0)|\Lambda_{Q}(v)\rangle & =\psi_{3s}(t_{1},t_{2})f^{(2)}u_{\gamma},\nonumber\\
v_{+}\langle0|[q_{1}^{T}(t_{1})C\gamma_{5}\bar{\slashed n}q_{2}(t_{2})]Q_{\gamma}(0)|\Lambda_{Q}(v)\rangle & =\psi_{4}(t_{1},t_{2})f^{(1)}u_{\gamma}.\label{LCDAdef}
\end{eqnarray}
The heavy quark field $Q$ is defined in the full QCD theory. In Ref.~\cite{Ball:2008fw} $Q$ should be denoted as $Q_{v}$ to stand for an effective field in HQET.  In this work, at the leading order we will not distinguish them. $\psi_{2}, \psi_{3\sigma},\psi_{3s}$ and $\psi_{4}$ are four LCDAs with different twists. $\gamma$  is a Dirac spinor index. $n$ and $\bar n$ are the two light-cone vectors, while $t_{i}$ are the distances between the $i$th light quark and the origin along the direction of $n$. The spacetime coordinate of the light quarks should be $t_{i}n^{\mu}$. The four-velocity of $\Lambda_{Q}$ is defined by light-cone coordinates $v^{\mu}=\frac{1}{2}(\frac{n^{\mu}}{v_{+}}+v_{+}\bar{n}^{\mu})$. In this work we simply choose the rest frame of $\Lambda_{Q}$, thus we have $v^{\mu}=\frac{1}{2}(n^{\mu}+\bar{n}^{\mu})$ and $v_+=1$. With the four LCDAs, one can express the matrix element $\epsilon_{abc}\langle\Lambda_{c}(v)|\bar{q}_{1k}^{a}(t_{1})\bar{q}_{2i}^{b}(t_{2})\bar{Q}_{\gamma}^{c}(0)|0\rangle$ as an expansion:
\begin{eqnarray}
\epsilon_{abc}\langle\Lambda_{c}(v)|\bar{q}_{1k}^{a}(t_{1})\bar{q}_{2i}^{b}(t_{2})\bar{Q}_{\gamma}^{c}(0)|0\rangle & =&\frac{1}{8}v_{+}\psi_{2}^{*}(t_{1},t_{2})f^{(1)}\bar{u}_{\gamma}(C^{-1}\gamma_{5}\bar{\slashed n})_{ki}\nonumber\\
 && -\frac{1}{8}\psi_{3\sigma}^{*}(t_{1},t_{2})f^{(2)}\bar{u}_{\gamma}(C^{-1}\gamma_{5}i\sigma^{\mu\nu})_{ki}\bar{n}_{\mu}n_{\nu}\nonumber\\
 && +\frac{1}{4}\psi_{3s}^{*}(t_{1},t_{2})f^{(2)}\bar{u}_{\gamma}(C^{-1}\gamma_{5})_{ki}\nonumber\\
 && +\frac{1}{8v_{+}}\psi_{4}^{*}(t_{1},t_{2})f^{(1)}\bar{u}_{\gamma}(C^{-1}\gamma_{5}\slashed n)_{kl},\label{quarksmatrix}
\end{eqnarray}
where we have explicitly shown the sum over color indexes $a,\ b,\ c$. The Fourier transformed form of the LCDAs are
\begin{eqnarray}
\psi(x_{1},x_{2})=\int_{0}^{\infty}d\omega_{1}d\omega_{2}e^{-i\omega_{1}t_{1}}e^{-i\omega_{2}t_{2}}\psi(\omega_{1},\omega_{2}),
\end{eqnarray}
where $\omega_{1}$ and $\omega_{2}$ are the momentum of the light
quarks along the light-cone. The total diquark momentum is defined as
$\omega=\omega_{1}+\omega_{2}$, and note that $x_{1}=t_{1}n$ , $x_{2}=t_{2}n$
\begin{eqnarray}
\psi(t_{1},t_{2}) &=&\int_{0}^{\infty}d\omega d\omega_{2}e^{-i\omega t_{1}}e^{-i\omega_{2}(t_{2}-t_{1})}\psi(\omega_{1},\omega_{2}), \\
\psi(0,t_{2})&=&\int_{0}^{\infty}d\omega\omega\int_{0}^{1}due^{-i\bar{u}\omega v\cdot x_{2}}\psi(\omega,u),
\end{eqnarray}
where $\omega_2=(1-u)\omega=\bar u \omega$. Here $t_{i}$ should be expressed in terms of Lorentz invariants $t_{i}=v\cdot x_{i}$.
Since in this work we will also consider the decays with $\Lambda_{c}$ in the final state, the LCDAs of $\Lambda_{c}$ are necessary. Although in the literatures there are no avaliable LCDAs of $\Lambda_{c}$,  due to heavy quark mass limit they are supposed to have the same form with those of $\Lambda_{b}$ given in Ref.~\cite{Ball:2008fw}. This argument can be trusted if one evaluate the energy of the light degree of freedom in $\Lambda_{Q}$ baryons: $m_{\Lambda_{Q}}-m_Q$. The ratio of such energies belonging to $\Lambda_{c}$ and $\Lambda_{b}$ respectively is almost one
\begin{eqnarray}
\frac{m_{\Lambda}-m_c}{m_{\Lambda_{b}}-m_b}=1.017,
\end{eqnarray}
where we choose $m_{\Lambda}=2.286$GeV, $m_{\Lambda_{b}}=5.62$GeV, $m_c=1.35$GeV, $m_b=4.7$GeV. Actually this is justified in HQET. Therefore, in this work we use the same LCDAs given in Ref.~\cite{Ball:2008fw} for both $\Lambda_{b}$ and $\Lambda_{c}$, which are expressed as
\begin{eqnarray}
 \psi_2(\omega,u) &=& \frac{15}{2} {\cal N}^{-1} \omega^2 \bar u u \int_{\omega/2}^{s_0} ds\, e^{-s/\tau} (s-\omega/2)\,,
\nonumber\\
 \psi_4(\omega,u) &=& 5 {\cal N}^{-1} \int_{\omega/2}^{s_0} ds\, e^{-s/\tau} (s-\omega/2)^3\,,
\nonumber\\
 \psi_{3s}(\omega,u) &=& \frac{15}{4} {\cal N}^{-1} \omega  \int_{\omega/2}^{s_0} ds\, e^{-s/\tau} (s-\omega/2)^2\,,
\nonumber\\
 \psi_{3\sigma}(\omega,u) &=& \frac{15}{4} {\cal N}^{-1} \omega (2u-1) \int_{\omega/2}^{s_0} ds\, e^{-s/\tau} (s-\omega/2)^2\,,
\label{SR:pert}
\end{eqnarray}
with 
\begin{eqnarray}
   {\cal N} &=& \int_0^{s_0}ds\, s^5 e^{-s/\tau}\,,
\end{eqnarray}
where $\tau$ and $s_0$ are the Borel parameter and the continuum threshold introduced by QCDSR in Ref.~\cite{Ball:2008fw}, which are taken to be in the interval 
$0.4 < \tau < 0.8$~GeV and a fixed value $s_0=1.2$~GeV respectively. Note that the LCDAs in Eq.~(\ref{SR:pert}) are only non-vanishing in the region $0<\omega <2s_0$.

\subsection{Light-Cone Sum Rules Framework}	

According to the framework of LCSR, to deal with the transition defined in Eq. (\ref{eq:parameterization2}), one needs to construct a two point correlation function
\begin{equation}
\Pi_{\mu}(p_{\Lambda},q) =i\int d^{4}xe^{iq\cdot x}\langle\Lambda_{Q^{\prime}}(p_{\Lambda})|T\{J^{V-A}_{\mu}(x)\bar{J}_{\Xi_{QQ^{\prime}}}(0)\}|0\rangle\label{eq:corrfunc}.
\end{equation}
The two currents $J^{V-A},\ J_{\Xi_{QQ^{\prime}}}$ are $V-A$ current and the $\Xi_{QQ^{\prime}}$ interpolating current respectively
\begin{equation}
J^{V-A}_{\mu}(x) =\bar{q}_{e}\gamma_{\mu}(1-\gamma_{5})Q_{e},
\end{equation}
while for $Q=Q^{\prime}=b,\ c$
\begin{equation}
J_{\Xi_{QQ}} =\epsilon_{abc}(Q_{a}^{T}C\gamma_{\mu}Q_{b})\gamma_{\mu}\gamma_{5}q^{\prime}_{c},
\end{equation}
for $Q=b,\ Q^{\prime}=c$
\begin{equation}
J_{\Xi_{bc}} =\frac{1}{\sqrt{2}}\epsilon_{abc}(b_{a}^{T}C\gamma^{\mu}c_{b}+c_{a}^{T}C\gamma^{\mu}b_{b})\gamma_{\mu}\gamma_{5}q^{\prime}_{c}.
\end{equation}

The correlation function Eq. (\ref{eq:corrfunc}) should be calculated both at hadron level and QCD level. At hadron level, by inserting a complete set of baryon states between $J^{V-A}$ and $J_{\Xi_{QQ^{\prime}}}$, and use the definition of $\Xi_{QQ^{\prime}}$ decay constant $f_{\Xi}$ 
\begin{eqnarray}
\langle {\Xi_{QQ^{\prime}}}(p_{ \Xi},s)|\bar{J}_{\Xi_{QQ^{\prime}}}(0)|0\rangle=f_{\Xi}\bar{u}_{\Xi}(p_{\Xi},s).
\end{eqnarray}
The correlation function induced by the vector current $\bar q \gamma^\mu Q$ can be expressed as
\begin{eqnarray}
\Pi_{\mu,V}^{hadron}(p_{\Lambda},q) & =&-\frac{f_{\Xi}}{(q+p_{\Lambda})^{2}-m_{\Xi}^{2}}\bar{u}_{\Lambda}(p_{\Lambda})[F_{1}(q^{2})\gamma_{\mu}+F_{2}(q^{2})p_{\Lambda\mu}+F_{2}(q^{2})p_{\Xi\mu}](\slashed q+\slashed p_{\Lambda}+m_{\Xi})+\ldots \nonumber\\
 &=&-\frac{f_{\Xi}}{(q+p_{\Lambda})^{2}-m_{\Xi}^{2}}\bar{u}_{\Lambda}(p_{\Lambda})\Big[F_{1}(q^{2})(m_{\Xi}-m_{\Lambda})\gamma_{\mu} \nonumber\\
 & &+[(m_{\Lambda}^{2}+m_{\Xi}m_{\Lambda})(F_{2}(q^{2})+F_{3}(q^{2}))+2m_{\Lambda}]v_{\mu} \nonumber\\
 & &+(m_{\Xi}+m_{\Lambda})F_{3}(q^{2})q_{\mu}+F_{1}(q^{2})\gamma_{\mu}\slashed q+m_{\Lambda}(F_{2}(q^{2})+F_{3}(q^{2}))v_{\mu}\slashed q+F_{3}(q^{2})q_{\mu}\slashed q\Big] \nonumber\\
 & &+\ldots,
\end{eqnarray}
where the ellipses stand for the contribution from continuum spectra $\rho^{h}$ above the threshold $s_{th}$, which has the integral form 
\begin{eqnarray}
\int_{s_{th}}^{\infty}ds\frac{\rho^{h}(s,q^{2})}{s-p_{\Xi}^{2}}.\label{correHadron}
\end{eqnarray}
For the correlation function induced by the axial-vector current $\bar q \gamma^\mu\gamma^5 Q$ the treatment is similar. In the following calculations we will mainly focus on the extraction of vector form factors $f_i$ while the the extraction of axial-vector form factors $g_i$ can be conducted analogously. 

Then we calculate the correlation function at QCD level. With the expansion of Eq.~(\ref{quarksmatrix}), the correlation function can be expressed as
\begin{eqnarray}
\Pi_{\mu,V}^{QCD}(p_{\Lambda},q) & =&-\frac{i}{4}\int d^{4}xe^{iq\cdot x}\{\psi_{2}^{*}(0,x)f^{(1)}\bar{u}[\gamma^{\nu}CS^{Q}(x)^{T}C^{T}\gamma_{\mu}\bar{\slashed n}\gamma_{\nu}] \nonumber\\
 && -\psi_{3\sigma}^{*}(0,x)f^{(2)}\bar{u}[\gamma^{\nu}CS^{Q}(x)^{T}C^{T}\gamma_{\mu}i\sigma^{\alpha\beta}\gamma_{\nu}]\bar{n}_{\alpha}n_{\beta} \nonumber\\
 && -2\psi_{3s}^{*}(0,x)f^{(2)}\bar{u}[\gamma^{\nu}CS^{Q}(x)^{T}C^{T}\gamma_{\mu}\gamma_{\nu}] \nonumber\\
 && +\psi_{4}^{*}(0,x)f^{(1)}\bar{u}[\gamma^{\nu}CS^{Q}(x)^{T}C^{T}\gamma_{\mu}\slashed n\gamma_{\nu}]\}.\label{Corre1}
\end{eqnarray}
It should be noted that the light-cone vectors $n$ and $\bar n$ in Eq.~(\ref{Corre1}) are chosen in a definite frame so that are not Lorentz covariant. They can be expressed in terms of Lorentz covariant form
\begin{equation}
n_{\mu} =\frac{1}{v\cdot x}x_{\mu},\ \ \ \bar{n}_{\mu}=2v_{\mu}-\frac{1}{v\cdot x}x_{\mu}.\label{nnbar}
\end{equation}
With the Fourier transformed LCDAs as well as light-cone vectors expressed in Eq.~\ref{nnbar}, the correlation function can be written as the form of convolution of diquark momenta $\omega$ and momenta fraction $u$
\begin{eqnarray}
\Pi_{\mu,V}^{QCD}(p_{\Lambda},q) & =&-\frac{i}{4}\int d^{4}x\int_{0}^{2s_0}d\omega\omega\int_{0}^{1}due^{i(q+\bar{u}\omega v)\cdot x}\nonumber\\
 && \times\{\psi_{2}(\omega,u)f^{(1)}\bar{u}_{\Lambda_{c}}[\gamma^{\nu}CS^{Q}(x)^{T}C^{T}\gamma_{\mu}(2\slashed v-\frac{\slashed x}{v\cdot x})\gamma_{\nu}]\nonumber\\
 && -\psi_{3\sigma}(\omega,u)f^{(2)}\bar{u}_{\Lambda_{c}}[\gamma^{\nu}CS^{Q}(x)^{T}C^{T}\gamma_{\mu}i\sigma^{\alpha\beta}\gamma_{\nu}]\frac{2v_{\alpha}x_{\beta}}{v\cdot x}\nonumber\\
 && -2\psi_{3s}(\omega,u)f^{(2)}\bar{u}_{\Lambda_{c}}[\gamma^{\nu}CS^{Q}(x)^{T}C^{T}\gamma_{\mu}\gamma_{\nu}]\nonumber\\
 && +\psi_{4}(\omega,u)f^{(1)}\bar{u}_{\Lambda_{c}}[\gamma^{\nu}CS^{Q}(x)^{T}C^{T}\gamma_{\mu}\frac{\slashed x}{v\cdot x}\gamma_{\nu}]\}.
\end{eqnarray}
Here $S^{Q}(x)$ is the usual free heavy quark propagator in QCD. After integrating the spacetime coordinate $x$, we can arrive at the explicit form of the correlation function at QCD level:
\begin{eqnarray}
&&\Pi_{\mu,V}^{QCD}((p_{\Lambda}+q)^2,q^2)\nonumber\\
 &=&\int_{0}^{2s_0}d\omega\omega\int_{0}^{1}du\psi_{2}(\omega,u)f^{(1)}\frac{1}{\frac{\bar{u}\omega}{m_{\Lambda}}s+H(u,\omega,q^{2})-m_{Q}^{2}}\bar{u}_{\Lambda_{c}}[-\bar{u}\omega\gamma_{\mu}+2(\bar{u}\omega+m_{Q})v_{\mu}+\gamma_{\mu}\slashed q]\nonumber\\
&+&\int_{0}^{2s_0}d\omega\int_{0}^{1}du\bar{u}f^{(1)}[\tilde{\psi}_{2}(\omega,u)-\tilde{\psi}_{4}(\omega,u)]\frac{1}{(\frac{\bar{u}\omega}{m_{\Lambda}}s+H(u,\omega,q^{2})-m_{Q}^{2})^{2}}\nonumber\\
 && \times\bar{u}_{\Lambda_{c}}\left[m_{Q}^{2}\gamma_{\mu}-2(m_{Q}+\bar{u}\omega)q_{\mu}-2\bar{u}\omega(m_{Q}+\bar{u}\omega)v_{\mu}-2q_{\mu}\slashed q-2\bar{u}\omega v_{\mu}\slashed q\right]\nonumber\\
 &+& 2\int_{0}^{2s_0}d\omega\int_{0}^{1}du\bar{u}\tilde{\psi}_{3\sigma}(\omega,u)f^{(2)}\frac{1}{(\frac{\bar{u}\omega}{m_{\Lambda}}s+H(u,\omega,q^{2})-m_{Q}^{2})^{2}}\bar{u}_{\Lambda_{c}}[-m_{Q}(q\cdot v)\gamma_{\mu}\nonumber\\
 && +2(m_{Q}+\bar{u}\omega+q\cdot v)q_{\mu}+(4\bar{u}\omega(q\cdot v)+q^{2}-3m_{Q}^{2}+3\bar{u}^{2}\omega^{2})v_{\mu}-m_{Q}\gamma_{\mu}\slashed q]\nonumber\\
 &+&\int_{0}^{2s_0}d\omega\omega\int_{0}^{1}du{\psi}_{3s}(\omega,u)f^{(2)}\frac{1}{\frac{\bar{u}\omega}{m_{\Lambda}}s+H(u,\omega,q^{2})-m_{Q}^{2}}\bar{u}_{\Lambda_{c}}[q_{\mu}+\bar{u}\omega v_{\mu}+\frac{1}{2}m_{Q}\gamma_{\mu}],\label{correQCD}
\end{eqnarray}
where $m_Q$ is the mass of the translating heavy quark, and
\begin{eqnarray}
&&s=(p_{\Lambda}+q)^2,\ \ 
q\cdot v=\frac{1}{2m_{\Lambda}}(s-q^{2}-m_{\Lambda}^{2}),\nonumber\\
&&H(u,\omega,q^{2})=\bar{u}\omega(\bar{u}\omega-m_{\Lambda})+(1-\frac{\bar{u}\omega}{m_{\Lambda}})q^{2}.
\end{eqnarray}
Here we have used the newly defined LCDAs
\begin{eqnarray}
\tilde{\psi}_i(\omega,u) =\int_{0}^{\omega}d\tau\tau\psi_i(\tau,u)\ \ \ \ (i=2,\ 3\sigma,\ 3s,\ 4).
\end{eqnarray}

\begin{figure}
\includegraphics[width=0.4\columnwidth]{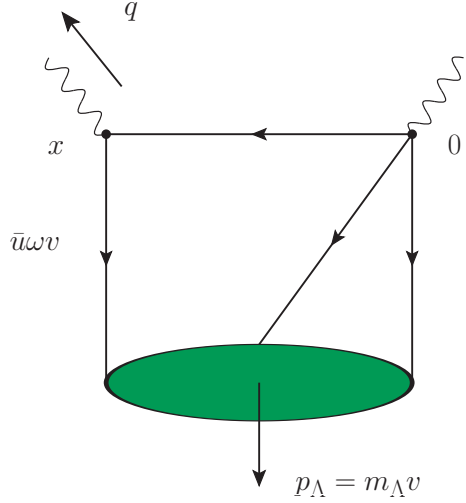} 
\caption{Feynman diagram of the QCD level correlation function. The green ellipse denotes the final $\Lambda_{Q^{\prime}}$ which has velocity $v$. The left black dot denotes the $V-A$ current while the right dot denotes the doubly-heavy baryon current. The left straight line denote one of the light quark inside the $\Lambda_{Q^{\prime}}$. It has momentum $\bar u \omega v$, where $\bar u$ is its momentum fraction related to the diquark. }
\label{fig:FeymDiag} 
\end{figure}

The Feynman diagram shown in Fig.~\ref{fig:FeymDiag} describes the correlation function at QCD level. Note that now the correlation function is expressed as a function of Lorentz invariants $(p_{\Lambda}+q)^2$ and $q^2$. By extracting the discontinuity of the correlation function Eq.~(\ref{correQCD}) acrossing the branch cut on the $(p_{\Lambda}+q)^2$ complex plane, one can write the correlation function as a dispersion integration form
\begin{eqnarray}
\Pi_{\mu,V}^{QCD}(p_{\Lambda},q)=\frac{1}{\pi}\int_{(m_{Q}+m_{Q^{\prime}}+m_{q})^{2}}^{\infty}ds\frac{\rm{Im}\Pi_{\mu,V}^{QCD}(s,q^2)}{s-(p_{\Lambda_{c}}+q)^{2}}.\label{DiscorreQCD}
\end{eqnarray}
According to the global Quark-Hadron duality, the integral  in
Eq.~(\ref{correHadron}) can be identified with the corresponding quantity
 at QCD level Eq.~(\ref{DiscorreQCD}). As a result, we have
\begin{eqnarray}
&&-\frac{f_{H}}{(q+p_{\Lambda})^{2}-m_{\Xi}^{2}}\bar{u}_{\Lambda}(p_{\Lambda})\Big[F_{1}(q^{2})(m_{\Xi}-m_{\Lambda})\gamma_{\mu}+[(m_{\Lambda}^{2}+m_{\Xi}m_{\Lambda})(F_{2}(q^{2})+F_{3}(q^{2}))+2m_{\Lambda}F_{1}(q^{2})]v_{\mu} \nonumber\\
 & &+(m_{\Xi}+m_{\Lambda})F_{3}(q^{2})q_{\mu}+F_{1}(q^{2})\gamma_{\mu}\slashed q+m_{\Lambda}(F_{2}(q^{2})+F_{3}(q^{2}))v_{\mu}\slashed q+F_{3}(q^{2})q_{\mu}\slashed q\Big]\nonumber\\
&&= \frac{1}{\pi}\int_{(m_{Q}+m_{Q^{\prime}}+m_{q})^{2}}^{sth}ds\frac{\rm{Im}\Pi_{\mu,V}^{QCD}(s,q^2)}{s-(p_{\Lambda_{c}}+q)^{2}}.\label{Formextract}
\end{eqnarray}
After constructing Borel transformation on the both sides of Eq.~(\ref{Formextract}), one can extract each of the form factors $F_i$. The $G_i$ can be obtained in a similar way.
Thus we obtain the explicit expression of each form factors
\begin{eqnarray}
F_{1}(q^{2}) & = & \frac{1}{f_{\Xi}(m_{\Xi}-m_{\Lambda})}exp(\frac{m_{\Xi}^{2}}{M^{2}})\nonumber \\
 & \times & \Big\{-\int_{0}^{1}du\int_{0}^{2s_{0}}d\omega\frac{m_{\Lambda}}{\bar{u}}exp(-\frac{s_{r}}{M^{2}})\theta(s_{th}-s_{r})\theta(s_{r}-(2m_{Q}+m_{q})^{2})\nonumber \\
 &  & \times[\psi_{3s}(\omega,u)f^{(2)}\frac{1}{2}m_{Q}-\psi_{2}(\omega,u)f^{(1)}\bar{u}\omega]\nonumber \\
 &  & +\int_{0}^{1}du\ \theta(\Delta)\theta(2s_{0}-\frac{\xi^{+}}{\bar{u}})\theta(\xi^{+})\frac{1}{\bar{u}\sqrt{\Delta}}\frac{m_{\Lambda}}{\omega}exp(-\frac{s_{0}}{M^{2}})\nonumber \\
 &  & \times\Big[(\tilde{\psi}_{2}(\omega,u)-\tilde{\psi}_{4}(\omega,u))f^{(1)}m_{Q}^{2}-\tilde{\psi}_{3\sigma}(\omega,u)f^{(2)}\frac{m_{Q}}{m_{\Lambda}}(s_{0}-q^{2}-m_{\Lambda}^{2})\Big]\Big|_{\omega=\frac{\xi^{+}}{\bar{u}}}\nonumber \\
 &  & -\int_{0}^{1}du\ \theta(\Delta)\theta(2s_{0}-\frac{\xi^{+}}{\bar{u}})\theta(\xi^{+})\frac{1}{\bar{u}\sqrt{\Delta}}\frac{m_{\Lambda}}{\omega}exp\Big(-\frac{(m_{Q}+m_{Q^{\prime}}+m_{q})^{2}}{M^{2}}\Big)\nonumber \\
 &  & \times\Big[(\tilde{\psi}_{2}(\omega,u)-\tilde{\psi}_{4}(\omega,u))f^{(1)}m_{Q}^{2}-\tilde{\psi}_{3\sigma}(\omega,u)f^{(2)}\frac{m_{Q}}{m_{\Lambda}}((m_{Q}+m_{Q^{\prime}}+m_{q})^{2}-q^{2}-m_{\Lambda}^{2})\Big]\Big|_{\omega=\frac{\xi^{+}}{\bar{u}}}\nonumber \\
 &  & -\int_{0}^{1}du\int_{0}^{\infty}d\omega\frac{m_{\Lambda}}{\bar{u}}exp(-\frac{s_{r}}{M^{2}})\theta(s_{th}-s_{r})\theta(s_{r}-(2m_{Q}+m_{q})^{2})\nonumber \\
 &  & \times\frac{d}{ds}\Big\{exp(-\frac{s}{M^{2}})\Big[(\tilde{\psi}_{2}(\omega,u)-\tilde{\psi}_{4}(\omega,u))f^{(1)}m_{Q}^{2}-\tilde{\psi}_{3\sigma}(\omega,u)f^{(2)}\frac{m_{Q}}{m_{\Lambda}}(s-q^{2}-m_{\Lambda}^{2})\Big]\Big\}\Big|_{s=s_{r}}\Big\},\nonumber \\
 ~\nonumber \\
F_{3}(q^{2}) & = & \frac{1}{f_{\Xi}(m_{\Xi}+m_{\Lambda})}exp(\frac{m_{\Xi}^{2}}{M^{2}})\nonumber \\
 & \times & \Big\{-\int_{0}^{1}du\int_{0}^{2s_{0}}d\omega\frac{m_{\Lambda}}{\bar{u}}exp(-\frac{s_{r}}{M^{2}})\theta(s_{th}-s_{r})\theta(s_{r}-(2m_{Q}+m_{q})^{2})\psi_{3s}(\omega,u)\nonumber \\
 &  & +\int_{0}^{1}du\ \theta(\Delta)\theta(2s_{0}-\frac{\xi^{+}}{\bar{u}})\theta(\xi^{+})\frac{1}{\bar{u}\sqrt{\Delta}}\frac{m_{\Lambda}}{\omega}exp(-\frac{s_{th}}{M^{2}})\nonumber \\
 &  & \times\Big[4(m_{Q}+\bar{u}\omega+\frac{s_{th}-q^2-m_{\Lambda}^{2}}{2m_{\Lambda}})\tilde{\psi}_{3\sigma}(\omega,u)f^{(2)}+2(m_{Q}+\bar{u}\omega)(\tilde{\psi}_{4}(\omega,u)-\tilde{\psi}_{2}(\omega,u))f^{(1)}\Big]\Big|_{\omega=\frac{\xi^{+}}{\bar{u}}}\nonumber \\
 &  & -\int_{0}^{1}du\ \theta(\Delta)\theta(2s_{0}-\frac{\xi^{+}}{\bar{u}})\theta(\xi^{+})\frac{1}{\bar{u}\sqrt{\Delta}}\frac{m_{\Lambda}}{\omega}exp\Big(-\frac{(m_{Q}+m_{Q^{\prime}}+m_{q})^{2}}{M^{2}}\Big)\nonumber \\
 &  & \times\Big[4\Big(m+\bar{u}\omega+\frac{(m_{Q}+m_{Q^{\prime}}+m_{q})^{2}-q^2-m_{\Lambda}^{2}}{2m_{\Lambda}}\Big)\tilde{\psi}_{3\sigma}(\omega,u)f^{(2)}\nonumber \\
 & &+2(m_{Q}+\bar{u}\omega)(\tilde{\psi}_{4}(\omega,u)-\tilde{\psi}_{2}(\omega,u))f^{(1)}\Big]\Big|_{\omega=\frac{\xi^{+}}{\bar{u}}}\nonumber \\
 &  & -\int_{0}^{1}du\int_{0}^{2s_{0}}d\omega\frac{m_{\Lambda}}{\bar{u}}exp(-\frac{s_{r}}{M^{2}})\theta(s_{th}-s_{r})\theta(s_{r}-(2m_{Q}+m_{q})^{2})\nonumber \\
 &  & \times\frac{d}{ds}\Big\{exp(-\frac{s}{M^{2}})\Big[4\Big(m_{Q}+\bar{u}\omega+\frac{s-q^2-m_{\Lambda}^{2}}{2m_{\Lambda}}\Big)\tilde{\psi}_{3\sigma}(\omega,u)f^{(2)}\nonumber \\
 & &+2(m_{Q}+\bar{u}\omega)(\tilde{\psi}_{4}(\omega,u)-\tilde{\psi}_{2}(\omega,u))f^{(1)}\Big]\Big\}\Big|_{s=s_{r}}\Big\},\nonumber \\
 ~\nonumber \\
F^{\prime}_{2}(q^{2}) & = & \frac{1}{f_{\Xi}}exp(\frac{m_{\Xi}^{2}}{M^{2}})\nonumber \\
 & \times & \Big\{-\int_{0}^{1}du\int_{0}^{2s_{0}}d\omega\frac{m_{\Lambda}}{\bar{u}}exp(-\frac{s_{r}}{M^{2}})\theta(s_{th}-s_{r})\theta(s_{r}-(2m_{Q}+m_{q})^{2})\nonumber \\
 &  & \times[2(\bar{u}\omega+m_{Q})\psi_{2}(\omega,u)f^{(1)}+\bar{u}\omega\psi_{3s}(\omega,u)f^{(2)}]\nonumber \\
 &  & +\int_{0}^{1}du\ \theta(\Delta)\theta(2s_{0}-\frac{\xi^{+}}{\bar{u}})\theta(\xi^{+})\frac{1}{\bar{u}\sqrt{\Delta}}\frac{m_{\Lambda}}{\omega}exp(-\frac{s_{th}}{M^{2}})\nonumber \\
 &  & \times\Big[2\bar{u}\omega(m_{Q}+\bar{u}\omega)(\tilde{\psi}_{4}(\omega,u)-\tilde{\psi}_{2}(\omega,u))f^{(1)}\nonumber \\
 &  &+2\tilde{\psi}_{3\sigma}(\omega,u)f^{(2)}(4\bar{u}\omega\frac{s_{th}-q^2-m_{\Lambda}^{2}}{2m_{\Lambda}}+q^{2}-3m_{Q}^{2}+3\bar{u}^{2}\omega^{2})\Big]\Big|_{\omega=\frac{\xi^{+}}{\bar{u}}}\nonumber \\
 &  & -\int_{0}^{1}du\ \theta(\Delta)\theta(2s_{0}-\frac{\xi^{+}}{\bar{u}})\theta(\xi^{+})\frac{1}{\bar{u}\sqrt{\Delta}}\frac{m_{\Lambda}}{\omega}exp\Big(-\frac{(m_{Q}+m_{Q^{\prime}}+m_{q})^{2}}{M^{2}}\Big)\nonumber \\
 &  & \times\Big[2\bar{u}\omega(m_{Q}+\bar{u}\omega)(\tilde{\psi}_{4}(\omega,u)-\tilde{\psi}_{2}(\omega,u))f^{(1)}\nonumber \\
 &  &+2\tilde{\psi}_{3\sigma}(\omega,u)f^{(2)}(4\bar{u}\omega\frac{(m_{Q}+m_{Q^{\prime}}+m_{q})^{2}-q^2-m_{\Lambda}^{2}}{2m_{\Lambda}}+q^{2}-3m_{Q}^{2}+3\bar{u}^{2}\omega^{2})\Big]\Big|_{\omega=\frac{\xi^{+}}{\bar{u}}}\nonumber \\
 &  & -\int_{0}^{1}du\int_{0}^{2s_{0}}d\omega\frac{m_{\Lambda}}{\bar{u}}exp(-\frac{s_{r}}{M^{2}})\theta(s_{th}-s_{r})\theta(s_{r}-(2m_{Q}+m_{q})^{2})\nonumber \\
 &  & \times\frac{d}{ds}\Big\{exp(-\frac{s}{M^{2}})\Big[2\bar{u}\omega(m_{Q}+\bar{u}\omega)(\tilde{\psi}_{4}(\omega,u)-\tilde{\psi}_{2}(\omega,u))f^{(1)}\nonumber \\
 &  &+2\tilde{\psi}_{3\sigma}(\omega,u)f^{(2)}\Big(4\bar{u}\omega\frac{s-q^2-m_{\Lambda}^{2}}{2m_{\Lambda}}+q^{2}-3m_{Q}^{2}+3\bar{u}^{2}\omega^{2}\Big)\Big]\Big\}\Big|_{s=s_{r}}\Big\},\nonumber \\
 ~\nonumber \\
F_{2}(q^{2}) & = & \frac{F_2^{\prime}(q^2)-2m_{\Lambda}F_1(q^2)}{m_{\Lambda}^2+m_{\Xi}m_{\Lambda}}-F_3(q^2),
\end{eqnarray}
where we have defined
\begin{eqnarray}
s_{r}& =&\frac{m_{\Lambda}}{\bar{u}\omega}(m_{Q}^{2}-H(u,\omega,q^{2})),\nonumber \\
\Delta & =&\frac{1}{m_{\Lambda_{c}^{2}}}(s_{th}-q^{2}-m_{\Lambda}^{2})-4(q^{2}-m_{Q}^{2}),\nonumber \\
\xi^{+} & =&\frac{1}{2}\left[-\frac{1}{m_{\Lambda}}(s_{th}-q^{2}-m_{\Lambda}^{2})+\sqrt{\Delta}\right].
\end{eqnarray}
For the axial-vector form factors, they are related with vector form factors as
\begin{eqnarray}
G_1(q^2)=F_1(q^2)\big|_{\psi_2\to-\psi_2,\ \psi_4\to-\psi_4}\nonumber\\
G_2(q^2)=F_2(q^2)\big|_{\psi_2\to-\psi_2,\ \psi_4\to-\psi_4}\nonumber\\
G_3(q^2)=F_3(q^2)\big|_{\psi_2\to-\psi_2,\ \psi_4\to-\psi_4}
\end{eqnarray}

From Eq.~(\ref{Formextract}), one could find that for each form factor there are two structures can be used to extract it. For example, for $f_1(q^2)$, one can extract it from both the $\gamma_{\mu}$ term and the $\gamma_{\mu}\slashed q$ term. However, only the $f_1(q^2)$ extracted from the $\gamma_{\mu}$ term can depend on all the four LCDAs. The criterion we follow here is to let all the four LCDAs contribute to each of the form factors. As a result, we extract the $f_1,\ f_2,\ f_3$ from the structures $\gamma_{\mu},\ v_{\mu},\ q_{\mu}$ respectively. Note that in Eq.~(\ref{Formextract}) the $v_{\mu}$ term contains all the three $f_i$ s, one needs to extract $f_1$ and $f_3$ firstly and then extract $f_3$ from the $v_{\mu}$ term.

\section{Numerical results}
\label{sec:numerical results}

\subsection{Transition Form Factors}

In this work, the heavy quark masses are taken as $m_{c}=(1.35\pm0.10)\ {\rm GeV}$
and $m_{b}=(4.7\pm0.1)\ {\rm GeV}$ while the masses of light quarks are approximated to zero. Tables~\ref{Tab:para_if} gives masses, lifetimes and decay constants $f_{\Xi}$ of doubly heavy baryons~\cite{Karliner:2014gca,Shah:2016vmd,Shah:2017liu,Kiselev:2001fw}. Decay constants of $\Lambda_Q$ defined in Eq.~(\ref{LCDAdef}) are taken as $f^{(1)}=f^{(2)}=0.03\pm0.005$, while the masses of $\Lambda_Q$ are taken as $m_{\Lambda_c}=2.286$ GeV and $m_{\Lambda_b}=5.620$ GeV. For the LCDA parameters in Eq.~(\ref{SR:pert}), we choose $s_0=1.2$ GeV and $\tau=(0.6\pm0.1$) GeV.

\begin{figure}
\includegraphics[width=0.9\columnwidth]{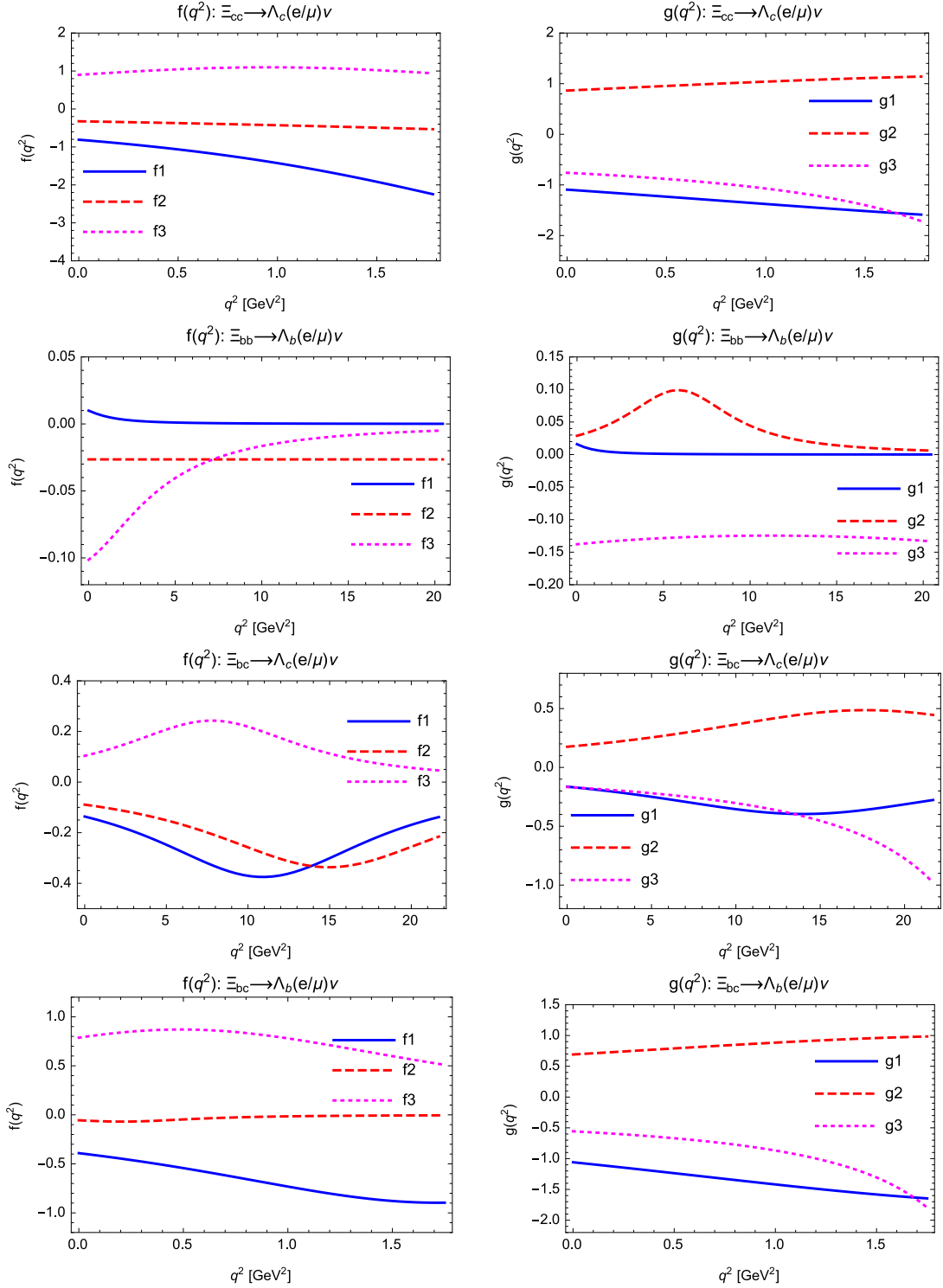} 
\caption{$q^2$ dependence of the $\Xi_{QQ^{\prime}}\to \Lambda_{Q^{\prime}}$ form factors. The first two graphs correspond to $\Xi_{cc}\to\Lambda_c$, the second two graphs correspond to $\Xi_{bb}\to\Lambda_b$, the third two graphs correspond to $\Xi_{bc}\to\Lambda_c$ and the fourth two graphs correspond to $\Xi_{bc}\to\Lambda_b$. Here the parameters $s_{th}$, $M^2$ are fixed at their center values as shown in Table~\ref{Tab:LCSRpars}, while $\tau=0.6$ GeV.}
\label{fig:Formfactors} 
\end{figure}

\begin{table}[!htb]
\caption{Masses, lifetimes and decay constants of doubly heavy baryons~\cite{Karliner:2014gca,Shah:2016vmd,Shah:2017liu,Kiselev:2001fw}.}
\label{Tab:para_if} %
\begin{tabular}{c|c|c|c}
\hline 
\hline
Baryons & Mass (GeV) & Lifetime (fs) &$f_{\Xi}$ $({\rm GeV}^{3})$\tabularnewline
\hline  
$\Xi_{cc}^{++}$  & $3.621$ \cite{Aaij:2017ueg}  & 256 & $0.109\pm0.020$ \tabularnewline
$\Xi_{bc}^{+}$  & $6.943$ \cite{Brown:2014ena}  & 244 & $0.150\pm0.035$ \tabularnewline
$\Xi_{bc}^{0}$  & $6.943$ \cite{Brown:2014ena}  & 93 & $0.150\pm0.035$ \tabularnewline
$\Xi_{bb}^{-}$  & $10.143$ \cite{Brown:2014ena}  & 370 & $0.199\pm0.052$ \tabularnewline
\hline 
\hline
\end{tabular}
\end{table}
The Borel parameters are chosen as to make the form factors be stable. The threshold $s_{th}$ of $\Xi_{QQ^{\prime}}$ and Borel parameters $M^2$ adopted in this work are shown in Table~\ref{Tab:LCSRpars}, which are consistent with those used in \cite{Shi:2019hbf}.  As argued by Ref.~\cite{Wang:2009hra}, the light-cone OPE for heavy baryon transition is expected to be reliable in the region where $q^2$ is positive but not too large. Thus the form factors need to be parametrized by a certain formula so as to be applicable at higher energy regions. The last column in Table~\ref{Tab:LCSRpars} lists the suitable $q^2$ regions for fitting the form factors. The numerical and fitting results for the form factors are given in Table~\ref{Tab:ff_cc_bb_bc}, where the results without asterisks are obtained by fitting the form factors with a double-pole parameterization function 
\begin{equation}
	F(q^{2})=\frac{F(0)}{1-\frac{q^{2}}{m_{{\rm fit}}^{2}}+\delta\left(\frac{q^{2}}{m_{{\rm fit}}^{2}}\right)^{2}}\label{eq:fit_formula_1},
\end{equation}
for the results with asterisks the above fitting function is slightly modified as
\begin{equation}
	F(q^{2})=\frac{F(0)}{1+\frac{q^{2}}{m_{{\rm fit}}^{2}}+\delta\left(\frac{q^{2}}{m_{{\rm fit}}^{2}}\right)^{2}}.\label{eq:fit_formula_2}
\end{equation}
For the form factors with weak $q^2$-dependence we will not parameterize them by the above two formulas. Here the form factor $f_{2}^{\Xi_{bb}\to\Lambda_{b}}$ is just kept as a constant equals to its value at $q^2=0$.  Since our theoretic calculation is based on LCSR, we would like to exam the exact error coming from the approach we used. Thus the error of the form factors are estimated from the thresholds $s_{th}$, Borel parameters $M^2$, and the LCDA parameter $\tau$, all of which characterize the framework of LCSR. The $q^2$ dependence of the form factors corresponding to the four channels are shown in Fig.~\ref{fig:Formfactors}, where the parameters $s_{th}$, $M^2$ are fixed at their center values as shown in Table~\ref{Tab:LCSRpars}, while $\tau=0.6$ GeV.

The comparison between this work and other works in the previous literatures are given in Table~\ref{Tab:comparison_cc} for the $\Xi_{cc}$ decays and Table~\ref{Tab:comparison_bb_bc} for the $\Xi_{bb}$ and $\Xi_{bc}$ decays. From the comparison one can find that most of the from factor obtained in this work are on the same order of magnitude as those of other works. Especially the results of $f_1(0)$ match well. However, our results of $g_1(0)$ are approximately an order of magnitude larger than those of other works, especially those from QCDSR~\cite{Shi:2019hbf} and LFQM~\cite{Wang:2017mqp}. On the other hand, the $f_1(0)$ s and the $g_1(0)$ s given in this work are at the same order.  As one know in the framework of HQET, both the form factors $f_1(0)$ and the $g_1(0)$ belonging to $B\to D$ transitions equal to the same Isgur-Wise function.  Although HQET cannot be applied for doubly heavy baryon decays, it seems that the effect of heavy quark symmetry still remains to some extent. 

\begin{table}
\caption{Threshold $s_{th}$ of $\Xi_{QQ^{\prime}}$, Borel parameters $M^2$, and $q^2$ range for fitting form factors.}
\label{Tab:LCSRpars}
\begin{center}
\begin{tabular}{c|c|c|c}
\hline 
\hline 
Channel & $s_{th}$ (GeV$^{2}$) & $M^{2}$ (GeV$^{2}$) & Fit Range (GeV$^{2}$)\tabularnewline
\hline 
$\Xi_{cc}\to\Lambda_{c}$ & $16\pm1$ & $6\pm1$ & $0<q^{2}<0.8$\tabularnewline
$\Xi_{bb}\to\Lambda_{b}$ & $112\pm2$ & $12\pm1$ & $0<q^{2}<3$\tabularnewline
$\Xi_{bc}\to\Lambda_{c}$ & $54\pm1.5$ & $9\pm1$ & $0<q^{2}<3$\tabularnewline
$\Xi_{bc}\to\Lambda_{b}$ & $54\pm1.5$ & $9\pm1$ & $0<q^{2}<0.8$\tabularnewline
\hline 
\hline 
\end{tabular}
\end{center}
\end{table}	
\begin{table}
\caption{The decay form factors of doubly-heavy baryons. $F(0)$, $m_{fit}$ and $\delta$ correspond to the three fitting parameters in Eq.~(\ref{eq:fit_formula_1}) or (\ref{eq:fit_formula_2}). The results without asterisks are obtained by fitting the form factors with Eq.~(\ref{eq:fit_formula_1}), while the results with asterisks are obtained by Eq.~(\ref{eq:fit_formula_2}).}
\label{Tab:ff_cc_bb_bc}
\begin{tabular}{c|c|c|c|c|c|c|c}
\hline 
\hline 
$F$  & $F(0)$  & $m_{{\rm fit}}$  & $\delta$  & $F$  & $F(0)$  & $m_{{\rm fit}}$  & $\delta$\tabularnewline
\hline 
$f_{1}^{\Xi_{cc}\to\Lambda_{c}}$  & $-0.81\pm0.01$  & $1.38\pm0.05$  & $0.34\pm0.01$  & $g_{1}^{\Xi_{cc}\to\Lambda_{c}}$  & $-1.09\pm0.02$  & $2.02\pm0.08$  & $0.66\pm0.05$ \tabularnewline
$f_{2}^{\Xi_{cc}\to\Lambda_{c}}$  & $-0.32\pm0.01$  & $1.92\pm0.08$  & $0.40\pm0.04$  & $g_{2}^{\Xi_{cc}\to\Lambda_{c}}$  & $0.86\pm0.02$  & $2.17\pm0.1$  & $0.95\pm0.11$ \tabularnewline
$f_{3}^{\Xi_{cc}\to\Lambda_{c}}$  & $0.9\pm0.07$  & $1.62\pm0.1$  & $1.38\pm0.7$  & $g_{3}^{\Xi_{cc}\to\Lambda_{c}}$  & $-0.76\pm0.01$  & $1.95\pm0.02$  & $-0.4\pm0.08$ \tabularnewline
\hline 
$f_{1}^{\Xi_{bb}\to\Lambda_{b}}$  & $-0.01\pm0.003^{*}$  & $1.33\pm0.24^{*}$  & $0.71\pm0.16^{*}$  & $g_{1}^{\Xi_{bb}\to\Lambda_{b}}$  & $-0.02\pm0.004^{*}$  & $1.1\pm0.13^{*}$  & $0.53\pm0.08^{*}$ \tabularnewline
$f_{2}^{\Xi_{bb}\to\Lambda_{b}}$  & $0.03\pm0.0$  & - -  & - -  & $g_{2}^{\Xi_{bb}\to\Lambda_{b}}$  & $-0.03\pm0.002$  & $2.03\pm0.04$  & $0.35\pm0.006$ \tabularnewline
$f_{3}^{\Xi_{bb}\to\Lambda_{b}}$  & $0.1\pm0.007^{*}$  & $3.34\pm0.13^{*}$  & $5.28\pm0.08^{*}$  & $g_{3}^{\Xi_{bb}\to\Lambda_{b}}$  & $0.14\pm0.003^{*}$  & $7.24\pm0.40^{*}$  & $-2.35\pm1.37^{*}$ \tabularnewline
\hline 
$f_{1}^{\Xi_{bc}\to\Lambda_{c}}$  & $-0.14\pm0.005$  & $2.93\pm0.06$  & $0.39\pm0.001$  & $g_{1}^{\Xi_{bc}\to\Lambda_{c}}$  & $-0.16\pm0.001$  & $3.45\pm0.05$  & $0.43\pm0.0$ \tabularnewline
$f_{2}^{\Xi_{bc}\to\Lambda_{c}}$  & $-0.09\pm0.002$  & $3.19\pm0.04$  & $0.34\pm0.001$  & $g_{2}^{\Xi_{bc}\to\Lambda_{c}}$  & $0.17\pm0.0$  & $3.72\pm0.04$  & $0.39\pm0.001$ \tabularnewline
$f_{3}^{\Xi_{bc}\to\Lambda_{c}}$  & $0.1\pm0.005$  & $2.6\pm0.08$  & $0.44\pm0.0$  & $g_{3}^{\Xi_{bc}\to\Lambda_{c}}$  & $-0.17\pm0.001$  & $4.43\pm0.03$  & $0.22\pm0.01$ \tabularnewline
\hline 
$f_{1}^{\Xi_{bc}\to\Lambda_{b}}$  & $0.39\pm0.01$  & $1.23\pm0.03$  & $0.44\pm0.02$  & $g_{1}^{\Xi_{bc}\to\Lambda_{b}}$  & $1.06\pm0.03$  & $1.77\pm0.06$  & $0.65\pm0.03$ \tabularnewline
$f_{2}^{\Xi_{bc}\to\Lambda_{b}}$  & $0.06\pm0.01$  & $0.73\pm0.03$  & $1.29\pm0.06$  & $g_{2}^{\Xi_{bc}\to\Lambda_{b}}$  & $-0.69\pm0.02$  & $1.89\pm0.06$  & $0.81\pm0.06$ \tabularnewline
$f_{3}^{\Xi_{bc}\to\Lambda_{b}}$  & $-0.79\pm0.06$  & $1.60\pm0.1$  & $2.62\pm1.15$  & $g_{3}^{\Xi_{bc}\to\Lambda_{b}}$  & $0.56\pm0.01$  & $1.79\pm0.01$  & $-0.48\pm0.04$ \tabularnewline
\hline 
\hline 
\end{tabular}
\end{table}

\begin{table}
\caption{Comparison of our results of $\Xi_{cc}$ decay form factors with the results from 
QCD sum rules (QCDSR)~\cite{Shi:2019hbf}, light-front quark model (LFQM)~\cite{Wang:2017mqp}, the nonrelativistic quark model (NRQM) and the MIT bag model (MBM)~\cite{PerezMarcial:1989yh}.}
\label{Tab:comparison_cc} \centering{}%
\begin{tabular}{c|c|c|c|c|c|c}
\hline
\hline 
Transitions  & $F(0)$  & This work  & QCDSR~\cite{Shi:2019hbf} & LFQM~\cite{Wang:2017mqp}  & NRQM ~\cite{PerezMarcial:1989yh}  & MBM ~\cite{PerezMarcial:1989yh} \tabularnewline
\hline 
$\Xi_{cc}^{++}\to\Lambda_{c}^{+}$  & $f_{1}(0)$  & $-0.81\pm0.01$  & $-0.59\pm0.05$  & $-0.79$  & $-0.36$  & $-0.45$\tabularnewline
 & $f_{2}(0)$  & $-0.32\pm0.01$  & $0.039\pm0.024$  & $0.008$  & $-0.14$  & $-0.01$\tabularnewline
 & $f_{3}(0)$  & $0.9\pm0.07$  & $0.35\pm0.11$  & - -  & $-0.08$  & $0.28$\tabularnewline
 & $g_{1}(0)$  & $-1.09\pm0.02$  & $-0.13\pm0.08$  & $-0.22$  & $-0.20$  & $-0.15$\tabularnewline
 & $g_{2}(0)$  & $0.86\pm0.02$  & $0.037\pm0.027$  & $0.05$  & $-0.01$  & $-0.01$\tabularnewline
 & $g_{3}(0)$  & $-0.76\pm0.01$  & $0.31\pm0.09$  & - -  & $0.03$  & $0.70$\tabularnewline
\hline 
\hline
\end{tabular}
\end{table}

\begin{table}
\caption{Comparison of our results on $\Xi_{bb}$ and $\Xi_{bc}$ decay form factors with the results from QCD sum rules (QCDSR)~\cite{Shi:2019hbf} and light-front quark model (LFQM)~\cite{Wang:2017mqp}.}
\label{Tab:comparison_bb_bc} \centering{}%
\begin{tabular}{c|c|c|c|c}
\hline 
\hline 
Transitions  & $F(0)$  & This work  & QCDSR~\cite{Shi:2019hbf} & LFQM~\cite{Wang:2017mqp} \tabularnewline
\hline 
$\Xi_{bb}\to\Lambda_{b}$  & $f_{1}(0)$  & $-0.01\pm0.003$  & $-0.086\pm0.013$  & $-0.102$ \tabularnewline
 & $f_{2}(0)$  & $0.03\pm0.0$  & $0.0022\pm0.0020$  & $0.0006$ \tabularnewline
 & $f_{3}(0)$  & $0.1\pm0.007$  & $0.0071\pm0.0072$  & - - \tabularnewline
 & $g_{1}(0)$  & $-0.02\pm0.004$  & $-0.074\pm0.013$  & $-0.036$ \tabularnewline
 & $g_{2}(0)$  & $-0.03\pm0.002$  & $0.0011\pm0.0024$  & $0.012$ \tabularnewline
 & $g_{3}(0)$  & $0.14\pm0.003$  & $0.0085\pm0.0055$  & - - \tabularnewline
\hline 
$\Xi_{bc}\to\Lambda_{b}$  & $f_{1}(0)$  & $0.39\pm0.01$  & $-0.65\pm0.06$  & $-0.55$ \tabularnewline
 & $f_{2}(0)$  & $0.06\pm0.01$  & $0.67\pm0.07$  & $0.30$ \tabularnewline
 & $f_{3}(0)$  & $-0.79\pm0.06$  & $-1.73\pm0.48$  & - - \tabularnewline
 & $g_{1}(0)$  & $1.06\pm0.03$  & $-0.15\pm0.08$  & $-0.15$ \tabularnewline
 & $g_{2}(0)$  & $-0.69\pm0.02$  & $-0.16\pm0.08$  & $0.10$ \tabularnewline
 & $g_{3}(0)$  & $0.56\pm0.01$  & $3.26\pm0.44$  & - - \tabularnewline
\hline 
$\Xi_{bc}\to\Lambda_{c}$  & $f_{1}(0)$  & $-0.14\pm0.005$  & $-0.11\pm0.01$  & $-0.11$ \tabularnewline
 & $f_{2}(0)$  & $-0.09\pm0.002$  & $-0.11\pm0.02$  & $-0.03$ \tabularnewline
 & $f_{3}(0)$  & $0.1\pm0.005$  & $0.16\pm0.03$  & - - \tabularnewline
 & $g_{1}(0)$  & $-0.16\pm0.001$  & $-0.085\pm0.014$  & $-0.047$ \tabularnewline
 & $g_{2}(0)$  & $0.17\pm0.0$  & $0.11\pm0.02$  & $0.02$ \tabularnewline
 & $g_{3}(0)$  & $-0.17\pm0.001$  & $-0.14\pm0.02$  & - - \tabularnewline
\hline 
\hline 
\end{tabular}
\end{table}
%%%%%%%%%%%%%%

%%%%%%%%%%%%%
\subsection{Semi-leptonic Decays}

In this section we consider the semi-leptonic decays of $\Xi_{QQ^{\prime}}\to\Lambda_{Q^{\prime}}$. The effective Hamiltonian inducing the semi-leptonic process is
\begin{eqnarray}
{\cal H}_{{\rm eff}} & = & \frac{G_{F}}{\sqrt{2}}\bigg(V_{ub}[\bar{u}\gamma_{\mu}(1-\gamma_{5})b][\bar{l}\gamma^{\mu}(1-\gamma_{5})\nu]+V_{cd}^{*}[\bar{d}\gamma_{\mu}(1-\gamma_{5})c][\bar{\nu}\gamma^{\mu}(1-\gamma_{5})l]\bigg),
\end{eqnarray}
where $G_{F}$ is Fermi constant and $V_{cs,cd,ub}$ are Cabibbo-Kobayashi-Maskawa
(CKM) matrix elements.

The decay amplitudes induced by vector current and axial-vector current are calculated with the use of helicity amplitudes respectively, they have the following expressions: 
\begin{eqnarray}
H_{\frac{1}{2},0}^{V} & = & -i\frac{\sqrt{Q_{-}}}{\sqrt{q^{2}}}\left((M_{1}+M_{2})f_{1}-\frac{q^{2}}{M_{1}}f_{2}\right),\;\;\;
H_{\frac{1}{2},0}^{A} =  -i\frac{\sqrt{Q_{+}}}{\sqrt{q^{2}}}\left((M_{1}-M_{2})g_{1}+\frac{q^{2}}{M}g_{2}\right),\nonumber \\
H_{\frac{1}{2},1}^{V} & = & i\sqrt{2Q_{-}}\left(-f_{1}+\frac{M_{1}+M_{2}}{M_{1}}f_{2}\right),\;\;\;
H_{\frac{1}{2},1}^{A}  =  i\sqrt{2Q_{+}}\left(-g_{1}-\frac{M_{1}-M_{2}}{M_{1}}g_{2}\right),\nonumber \\
H_{\frac{1}{2},t}^{V} & = & -i\frac{\sqrt{Q_{+}}}{\sqrt{q^{2}}}\left((M_{1}-M_{2})f_{1}+\frac{q^{2}}{M_{1}}f_{3}\right),\;\;\;
H_{\frac{1}{2},t}^{A} =  -i\frac{\sqrt{Q_{-}}}{\sqrt{q^{2}}}\left((M_{1}+M_{2})g_{1}-\frac{q^{2}}{M_{1}}g_{3}\right),
\end{eqnarray}
where $Q_{\pm}=(M_1\pm M_2)^{2}-q^{2}$ and $M_{1}(M_{2})$ is the mass of the initial (final) baryon. The amplitudes with negative helicity are related to those with positive helicity
\begin{equation}
H_{-\lambda_{2},-\lambda_{W}}^{V}=H_{\lambda_{2},\lambda_{W}}^{V}\quad\text{and}\quad H_{-\lambda_{2},-\lambda_{W}}^{A}=-H_{\lambda_{2},\lambda_{W}}^{A},
\end{equation}
where the polarizations of the final $\Lambda_{Q^{\prime}}$ and the intermediate $W$ boson are denoted by $\lambda_{2}$ and $\lambda_{W}$, respectively. The total helicity amplitudes induced by the $V-A$ current are  
\begin{equation}
H_{\lambda_{2},\lambda_{W}}=H_{\lambda_{2},\lambda_{W}}^{V}-H_{\lambda_{2},\lambda_{W}}^{A}.
\end{equation}

\begin{figure}
\includegraphics[width=0.9\columnwidth]{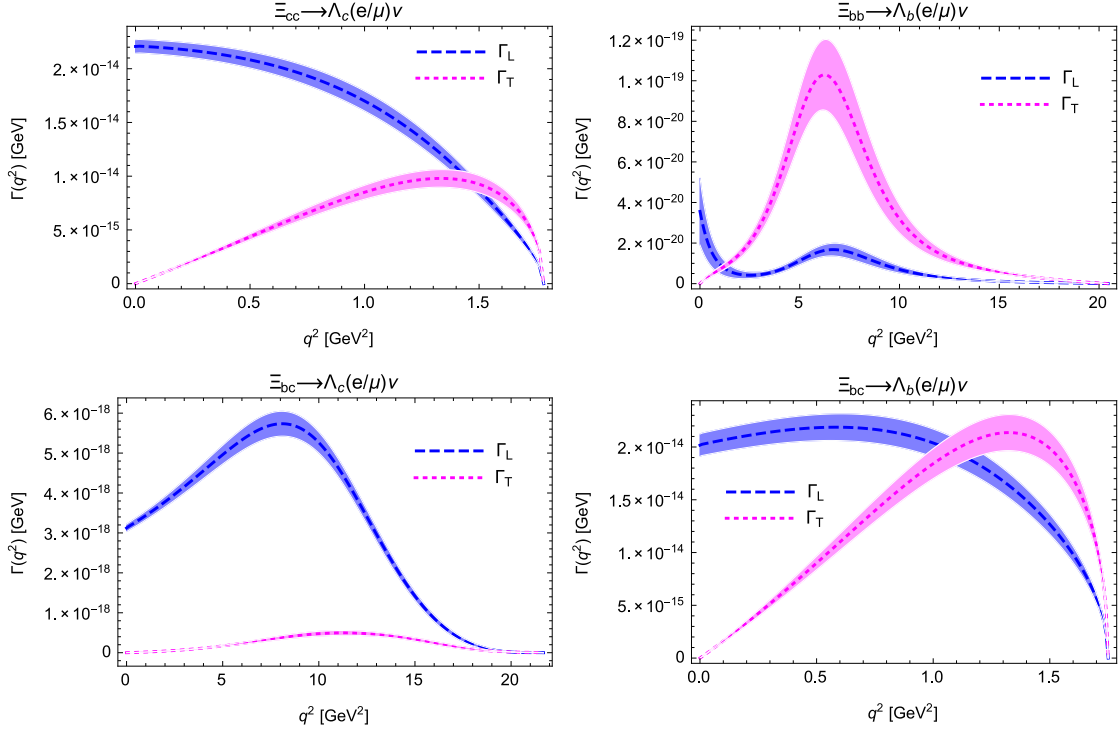} 
\caption{$q^2$ dependence of the semi-leptonic $\Xi_{QQ^{\prime}}\to\Lambda_{Q^{\prime}}l\nu_l$ decay widths. The blue bands correspond to $\Gamma_L$ while the red bands correspond to $\Gamma_T$. The dashed lines describe the center value curves and the band width reflects the error.}
\label{fig:SemiDecayWidths} 
\end{figure}

Decay widths of $\Xi_{QQ^{\prime}}\to\Lambda_{Q^{\prime}}l\nu$ can be separated into two parts which correspond to the longitudinally and transversely polarized $l\nu$ pairs respectively
\begin{align}
\frac{d\Gamma_{L}}{dq^{2}} & =\frac{G_{F}^{2}|V_{{\rm CKM}}|^{2}q^{2}\ p\ (1-\hat{m}_{l}^{2})^{2}}{384\pi^{3}M_{1}^{2}}\left((2+\hat{m}_{l}^{2})(|H_{-\frac{1}{2},0}|^{2}+|H_{\frac{1}{2},0}|^{2})+3\hat{m}_{l}^{2}(|H_{-\frac{1}{2},t}|^{2}+|H_{\frac{1}{2},t}|^{2})\right),\label{eq:longi-1}\\
\frac{d\Gamma_{T}}{dq^{2}} & =\frac{G_{F}^{2}|V_{{\rm CKM}}|^{2}q^{2}\ p\ (1-\hat{m}_{l}^{2})^{2}(2+\hat{m}_{l}^{2})}{384\pi^{3}M_{1}^{2}}(|H_{\frac{1}{2},1}|^{2}+|H_{-\frac{1}{2},-1}|^{2}),\label{eq:trans-1}
\end{align}
%\begin{align}
%\frac{d\Gamma_{L}}{dq^{2}} & =\frac{G_{F}^{2}|V_{{\rm CKM}}|^{2}q^{2}\ p\ (1-\hat{m}_{l}^{2})^{2}}{384\pi^{3}M_{1}^{2}}\left((2+\hat{m}_{l}^{2})(|H_{-\frac{1}{2},0}|^{2}+|H_{\frac{1}{2},0}|^{2})+3\hat{m}_{l}^{2}(|H_{-\frac{1}{2},t}|^{2}+|H_{\frac{1}{2},t}|^{2})\right),\label{eq:longi-1}\\
%\frac{d\Gamma_{T}}{dq^{2}} & =\frac{G_{F}^{2}|V_{{\rm CKM}}|^{2}q^{2}\ p}{192\pi^{3}M_{1}^{2}}(|H_{\frac{1}{2},1}|^{2}+|H_{-\frac{1}{2},-1}|^{2}),\label{eq:trans-1}
%\end{align}
where $\hat{m}_{l}\equiv m_{l}/\sqrt{q^{2}}$, $p=\sqrt{Q_{+}Q_{-}}/(2M_{1})$
is the three-momentum magnitude of $\Lambda_{Q^{\prime}}$ in the rest frame of $\Xi_{QQ^{\prime}}$. Here the Fermi constant and CKM matrix elements are taken from~\cite{Olive:2016xmw,Tanabashi:2018oca}: 
\begin{align}
&G_{F}=1.166\times10^{-5}{\rm GeV}^{-2},\nonumber \\
&|V_{ub}|=0.00357,\quad|V_{cd}|=0.225.\label{eq:GFCKM}
\end{align}
By integrating out the squared transfer momentum $q^{2}$, one can obtain the total decay width 
\begin{equation}
\Gamma=\int_{m_l^2}^{(M_{1}-M_{2})^{2}}dq^{2}\frac{d\Gamma}{dq^{2}},
\end{equation}
where
\begin{equation}
\frac{d\Gamma}{dq^{2}}=\frac{d\Gamma_{L}}{dq^{2}}+\frac{d\Gamma_{T}}{dq^{2}}.
\end{equation}
Table~\ref{Tab:semi_lep} shows the integrated partial decay widths, branching ratios and the ratios of $\Gamma_{L}/\Gamma_{T}$ for various semi-leptonic $\Xi_{QQ^{\prime}}\to\Lambda_{Q^{\prime}}l(\tau)\nu_l$ processes, where $l=e/\mu$. The masses of $e$ and $\mu$ are approximated to zero while the mass of $\tau$ is taken as $1.78$ GeV~\cite{Olive:2016xmw}. Fig.~\ref{fig:SemiDecayWidths} shows the $q^2$ dependence of the differential decay widths corresponding to four channels. Table~\ref{Tab:comparison_semi_lep} gives a comparison of our decay width results with those given in the literatures.

\begin{table}
\caption{Decay widths and branching ratios of the semi-leptonic $\Xi_{QQ^{\prime}}\to\Lambda_{Q^{\prime}}l\nu_l$ decays, where $l=e/\mu$.}
\label{Tab:semi_lep}%
\begin{tabular}{l|c|c|c}
\hline 
\hline 
Channels  & $\Gamma/{\rm GeV}$  & $\cal{B}$  & $\Gamma_{L}/\Gamma_{T}$ \tabularnewline
\hline
$\Xi_{cc}^{++}\to\Lambda_{c}^{+}l^{+}\nu_{l}$  & $(3.95\pm0.21)\times10^{-14}$ & $(1.53\pm0.1)\times10^{-2}$ & $2.6\pm0.35$ \tabularnewline
$\Xi_{bb}^{-}\to\Lambda_{b}^{0}l^{-}\nu_{l}$  & $(7.35\pm1.43)\times10^{-19}$ & $(4.13\pm0.8)\times10^{-7}$ & $0.21\pm0.12$\tabularnewline
$\Xi_{bb}^{-}\to\Lambda_{b}^{0}\tau^{-}\nu_{l}$  & $(6.1\pm1.1)\times10^{-19}$ & $(3.43\pm0.65)\times10^{-7}$ & $0.08\pm0.04$\tabularnewline
$\Xi_{bc}^{0}\to\Lambda_{c}^{+}l^{-}\nu_{l}$  & $(7.17\pm0.4)\times10^{-17}$ & $(1.01\pm0.06)\times10^{-5}$ & $13.38\pm2.74$\tabularnewline
$\Xi_{bc}^{0}\to\Lambda_{c}^{+}\tau^{-}\nu_{l}$  & $(4.09\pm0.28)\times10^{-17}$ & $(5.77\pm0.4)\times10^{-6}$ & $7.38\pm1.61$\tabularnewline
$\Xi_{bc}^{+}\to\Lambda_{b}^{0}l^{+}\nu_{l}$  & $(5.51\pm0.38)\times10^{-14}$ & $(2.04\pm0.14)\times10^{-2}$ & $1.39\pm0.21$ \tabularnewline
\hline 
\hline 
\end{tabular}
\end{table}
\begin{table}
\caption{Comparison of the decay widths (in units of GeV) for
the semi-leptonic decays in this work with the results derived from QCD sum rules (QCDSR)~\cite{Shi:2019hbf}, the light-front quark model (LFQM)~\cite{Wang:2017mqp}, the heavy quark spin symmetry (HQSS)~\cite{Albertus:2012nd}, the nonrelativistic
quark model (NRQM)~\cite{PerezMarcial:1989yh} and the MIT bag model (MBM)~\cite{PerezMarcial:1989yh} in literatures.}
\label{Tab:comparison_semi_lep} \centering{}%
\resizebox{\textwidth}{15mm}{
\begin{tabular}{c|c|c|c|c|c|c}
\hline 
\hline 
Channels  & This work  & QCDSR~\cite{Shi:2019hbf} & LFQM~\cite{Wang:2017mqp}  & HQSS~\cite{Albertus:2012nd}  & NRQM~\cite{PerezMarcial:1989yh}  & MBM~\cite{PerezMarcial:1989yh}\tabularnewline
\hline 
$\Xi_{cc}^{++}\to\Lambda_{c}^{+}l^{+}\nu_{l}$  & $(3.95\pm0.21)\times10^{-14}$  & $(6.1\pm1.1)\times10^{-15}$  & $1.05\times10^{-14}$  & $3.20\times10^{-15}$  & $1.97\times10^{-15}$  & $1.32\times10^{-15}$\tabularnewline
\hline 
$\Xi_{bb}^{-}\to\Lambda_{b}^{0}l^{-}\bar{\nu}_{l}$  & $(7.35\pm1.43)\times10^{-19}$  & $(3.0\pm0.7)\times10^{-17}$  & $1.58\times10^{-17}$  & - -  & - -  & - -\tabularnewline
\hline 
$\Xi_{bc}^{0}\to\Lambda_{c}^{+}l^{-}\bar{\nu}_{l}$  & $(7.17\pm0.4)\times10^{-17}$  & $(2.2\pm0.5)\times10^{-17}$  & $1.84\times10^{-17}$  & - -  & - -  & - -\tabularnewline
\hline 
$\Xi_{bc}^{+}\to\Lambda_{b}^{0}l^{+}\nu_{l}$  & $(5.51\pm0.38)\times10^{-14}$  & $(1.1\pm0.2)\times10^{-14}$  & $6.85\times10^{-15}$  & - -  & - -  & - -\tabularnewline
\hline 
\hline 
\end{tabular}}
\end{table}

There are several remarks:
\begin{itemize}
\item The error of the decay widths given in Table~\ref{Tab:semi_lep} and Fig.~\ref{fig:SemiDecayWidths} both come from the error of form factors.

\item From Table~\ref{Tab:semi_lep}, one can find that the decay widths and branching ratios of $c\to d$ processes are several orders of magnitude larger than those of $b\to u$ processes. This feature is compatible with the case of $B$ and $D$ decays.

\item According to the SU(3) symmetry, the decay widths of various semi-leptonic channels are related with each other. Ref.\cite{Wang:2017azm,Shi:2017dto} have offered a systematic SU(3) analysis of doubly heavy baryon decays as well as a complete decay width relations. Although in this work only the processes with $\Lambda_{Q^{\prime}}$ final states are considered, one can still estimate decay widths of several other channels from Ref.\cite{Wang:2017azm}:
\begin{eqnarray}
\Gamma(\Omega_{cc}^{+}\to\Xi_{c}^{0}l^{+}\nu)&=&\Gamma(\Xi_{cc}^{++}\to\Lambda_{c}^{+}l^{+}\nu)=(3.95\pm0.21)\times10^{-14} \rm{GeV},\nonumber\\
\Gamma(\Omega_{bc}^{0}\to\Xi_{b}^{-}l^{+}\nu)&=&\Gamma(\Xi_{bc}^{+}\to\Lambda_{b}^{0}l^{+}\nu)=(5.51\pm0.38)\times10^{-14} \rm{GeV},\nonumber\\
\Gamma(\Omega_{bb}^{-}\to\Xi_{b}^{0}l^{-}\bar{\nu})&=&\Gamma(\Xi_{bb}^{-}\to\Lambda_{b}^{0}l^{-}\bar{\nu})=(7.35\pm1.43)\times10^{-19} \rm{GeV}.
\end{eqnarray}

\item From the comparison shown in Table~\ref{Tab:comparison_semi_lep}, it seems that the  semi-leptonic decay widths derived in this and other works are approximately on the same order of magnitude.

\end{itemize}

\section{Conclusions}

In summary, we have presented a study on the semi-leptonic decay of doubly heavy baryons into an anti-triplet baryon $\Lambda_Q$. We derived the baryon transition form factors with LCSR, where the LCDAs of $\Lambda_b$ are used for both $\Lambda_b$ and $\Lambda_c$ final states due to the heavy quark symmetry. From the numerical results of our form factors, we find that $f_1$ and $g_1$ are at the same magnitude order, which seems consistent with HQET. The obtained form factors are then used for predicting the semi-leptonic doubly-heavy baryon decay widths as well as the branching ratios. Most of them are consistent with the phenomenology results given in other works. We hope our use of LCSR for double-heavy baryon transitions can help us test or even understand the light-cone dynamics of heavy baryon states, while the phenomenology predictions given in this work can be tested by future measurement by LHCb as well as other experiments.

\label{sec:conclusions}

\section*{Acknowledgements}
The authors are very grateful to Profs. Wei Wang and Yu-Ming Wang for useful discussions. This work is supported in part by National
Natural Science Foundation of China under Grants No.11575110,  
11735010, Natural Science Foundation of Shanghai under Grants
No.~15DZ2272100 and No.~15ZR1423100, and by Key Laboratory for Particle
Physics, Astrophysics and Cosmology, Ministry of Education.

\end{document}